\documentclass[aps,pra,superscriptaddress,twocolumn,citeautoscript,10pt, rsi,reprint]{revtex4-1}
\usepackage{tabularx}  
\usepackage{amsmath}
\usepackage{graphicx}
\usepackage{bm} 
\usepackage{soul}
\usepackage{xcolor}
\usepackage{amsmath}
\usepackage{float}
\usepackage{comment}
\usepackage{graphicx}
\usepackage{dcolumn}
\usepackage{multirow}
\usepackage{booktabs}
\usepackage{mhchem}
\usepackage{xr}
\usepackage[colorlinks=true,bookmarks=false,citecolor=blue,linkcolor=blue,hyperfootnotes=true,urlcolor=blue]{hyperref}

\def\CHR{Cr$_2$O$_3$}  
\def\NEL{N\'eel}
\def\NELT{$T_{\mathrm{N}}$}

\begin{document}
\preprint{AIP/123-QED}
\title{Intrinsic defects and 4$d$/5$d$ transition metal defects in Cr$_2$O$_3$: pathways for enhancing the N\'eel temperature}
\author{Xuecong Wang}
\author{Sai Mu}\email{mus@mailbox.sc.edu}
\affiliation{Department of Physics and Astronomy and SmartState Center for Experimental Nanoscale Physics, University of South Carolina, Columbia, South Carolina, 29208, USA}

\begin{abstract}
First-principles calculations are employed to explore avenues to increase the \NEL\ temperature ($T_{\mathrm{N}}$) of the magnetoelectric antiferromagnet \CHR\ through doping. Employing the hybrid functional method, we calculate the formation energy of intrinsic defects and transition metal dopants (Mo, W, Nb, Ta, Zr, and Hf) to assess their likelihood of formation. 
Intrinsic defect calculations indicate that Cr interstitials and oxygen vacancies dominate under Cr-rich conditions, whereas Cr vacancies prevail under O-rich conditions. Notably, under Cr-rich conditions, the Fermi level can be pinned slightly above mid-gap due to the formation of Cr interstitials and oxygen vacancies. Additionally, we identify a split Cr vacancy configuration that is more energetically stable than a normal Cr vacancy in corundum \CHR.
To assess the influence of dopants on \NELT\ of \CHR, we calculate the enhancement of the exchange energy for the spin at the dopant site or at an adjacent Cr site using the supercell method.Our study identifies isovalent Mo and W substitutions at Cr sites as the most promising candidates for increasing the \NEL\ temperature because of impurity-mediated enhancement of the exchange interaction for half-filled bands. Formation-energy calculations indicate that Mo and W substitutions at Cr sites form more readily under Cr-rich conditions and that a Fermi level near or slightly above midgap stabilizes the desirable \textit{neutral} charge states of the Mo and W defects. This is assisted by the formation of intrinsic Cr interstitials and O vacancies under Cr-rich conditions.
These findings offer a route to utilize defects for higher \NELT\ and enhanced performance of \CHR\ in magnetoelectric devices and provide useful guidance for future experimental studies.

\end{abstract}
\maketitle
\section{\label{sec:level1}Introduction}

Magnetoelectric antiferromagnets~\cite{landauelectrodynamics} have garnered considerable attention due to their applications in magnetoelectronic devices that leverage the electric control of magnetization~\cite{binek2004magnetoelectronics,belashchenko2010equilibrium,he2010robust}. Among these materials, \CHR\ serves as the active magnetoelectric material in voltage-controlled exchange bias devices, which are promising for magnetic memory and logic applications because of their nonvolatility and low power consumption.\cite{chen2006magnetoelectric}

While \CHR\ exhibits the highest \NEL\ temperature (\NELT\ = $307 K$) among magnetoelectric antiferromagnets~\cite{borovik-romanov2006international}, its \NELT, which is near room temperature, is still insufficiently high for practical applications. This motivates further investigation aimed at increasing the \NELT\ of \CHR. 
Two prominent approaches have been explored to achieve higher \NELT: mechanical strain~\cite{mu2014first,kota2013strain} and chemical doping~\cite{mu2013effect,street2014increasing}. 
In this study, we concentrate on the doping approach. 
 A previous first-principles study suggested that substitutional boron doping on the oxygen site could effectively raise the \NELT\ of \CHR~through impurity-mediated superexchange-like interactions \cite{mu2013effect}, a finding that was later demonstrated experimentally~\cite{street2014increasing,mahmood2021voltage}.  
As opposed to the anion impurity, cation impurities have been less successful in achieving high \NELT\ of \CHR. Incorporation of 3$d$ transition metal dopants (Ti, V, Mn, Fe, Co, Ni)~\cite{mu2013effect} and \textit{isovalent} 4$d$ transition metal dopants (Zr, Nb, Mo)~\cite{mu2019influence} on the cation Cr site has been investigated for a higher \NELT. However, \emph{only} Mo exhibits some potential for a higher \NELT~\cite{mu2019influence}, while the remaining candidates have proven detrimental to \NELT\ enhancement. 

Previous calculations on enhancing the \NELT\ of \CHR~have focused exclusively on substitutional dopants, neglecting the possible incorporation of defects at interstitial sites. Additionally, the likelihood of incorporation of these defects remains unexplored.  
Furthermore, previous first-principles studies have only considered isovalent defects, overlooking other possible charge states of the dopant and their impact on \NELT. Consequently, it is important to perform defect-formation-energy calculations to evaluate the likelihood of defect incorporation, determine the feasible charge states of the defect, and assess their influence on the \NELT.

Intrinsic defects determine the electrical properties of \CHR, with the predominant ones significantly influencing the position of the Fermi level and consequently the charge state of extrinsic defects. Moreover, the spin states of these intrinsic defects might affect their formation energies and the magnetism of \CHR.
Although the formation energies of some of the intrinsic defects have been explored previously~\cite{medasani2017vacancies,medasani2018first,medasani2019temperature}, the accuracy of previous results is limited by the methods used to account for self-interaction errors, and the spin states of intrinsic defects have not been addressed clearly in these studies.

Using density functional theory calculations, we conduct a comprehensive investigation of formation energies of both intrinsic defects and selected 4$d$ and 5$d$ transition metal foreign defects. The likelihood of defect formation at cation and interstitial sites, as well as the accessible charge states, is evaluated using hybrid-functional calculations. The influence of the defects on the \NELT\ of \CHR~is explored, and we find that neutral Mo and W defects are the most promising for raising the \NELT\ and that Cr-rich conditions facilitate their formation. A rough estimate indicates that 1~\% Mo (W) doping increases the \NELT\ by 5.8~\% (8.9~\%).
Intrinsic defect calculations further demonstrated that the neutral charge state of Mo could be realized under Cr-rich conditions, through the Fermi-level pinning by the predominant Cr interstitial and O vacancy.

\section{Methodology} \label{sec:method}

\subsection{Computational details}\label{sec:calc_detail}

\begin{figure}
\includegraphics[width=0.47\textwidth]{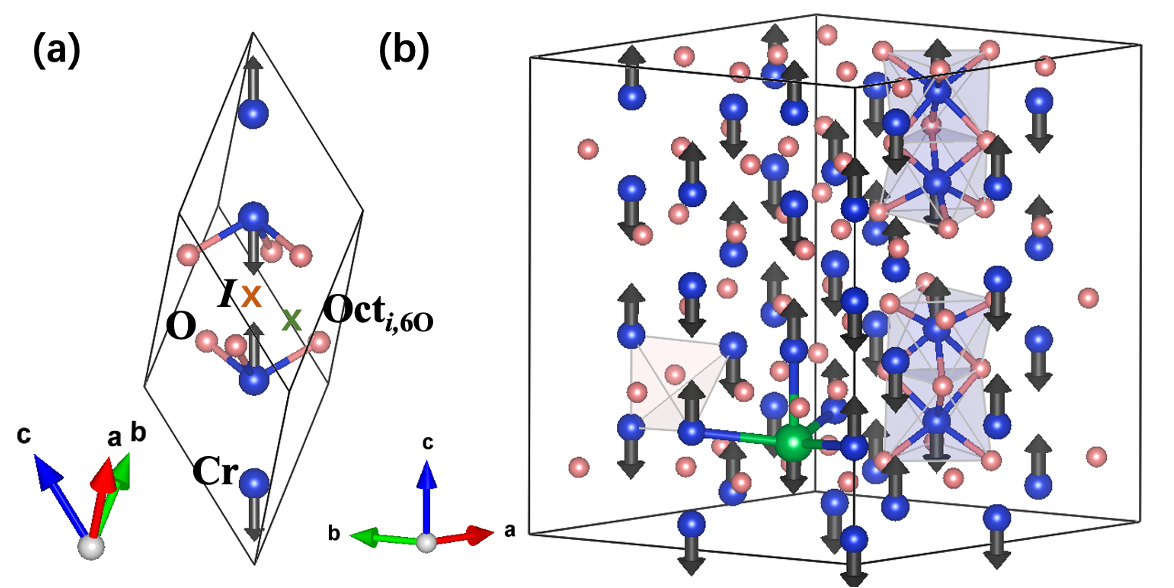}
\caption{\label{fig:structure} (Color online) (a) Rhombohedral primitive cell of \CHR. Larger blue and pink spheres represent Cr atoms with opposite spin moments; smaller red spheres display the O atoms.  (b) $2 ~ \times  ~ 2 ~ \times ~ 1$ hexagonal supercell with doped Mo atom (green sphere). The bonds connect Mo defect with the nearest Cr neighbour and three second nearest Cr neighbor. }
\end{figure}

Our first-principles calculations were performed using the Vienna Ab initio Simulation Package (VASP 6.3.2)~\cite{kresse1996efficient,kresse1996efficiency}. Projector augmented wave potentials~\cite{blochl1994projector} were used and exchange and correlation were treated within the generalized gradient approximation (GGA)~\cite{godby1986accurateGGA}. The valence electron wavefunctions were expanded with a kinetic energy cutoff of 520 eV and the valence-electron configurations were 3$d^5$4$s^1$ for Cr and  2$s^2$2$p^4$ for O. We simulated bulk \CHR\ using the 10-atom rhombohedral primitive cell [see Fig.~\ref{fig:structure} (a)]. 
Four Cr atoms contain two different sublattices with opposite spins, forming $+-+-$ antiferromagnetic spin ordering. Gaussian smearing of 0.1 eV (0.02 eV) and a $\mathit{\Gamma}$-centered 3 × 3 × 3 (5 × 5 × 5) Monkhorst-Pack $\mathit{k}$-point mesh~\cite{monkhorst1976specialkpoint} were used for the structural relaxation [density of states (DOS) calculation].  

To account for Coulomb correlations of the localized $d$ orbitals on Cr and on transition-metal defects, we employed the hybrid functional of Heyd, Scuseria, and Ernzerhof (HSE)~\cite{heyd2003hybrid} with a mixing parameter $\alpha = 0.20$ and a screening parameter of $0.3$. The combination of these parameters yields a band gap of 3.45 eV, which agrees well with the experimental value (3.4 eV) \cite{adler1968insulating,crawford1964electricalCr2O3,zaanen1985bandgap}. 
The effects of the mixing and screening parameters on the band gap and lattice parameters were investigated, and the optimal parameter combination was selected accordingly. More details can be found in Supplemental Materials (SM)~\cite{supplemental}.

Defects were simulated using the supercell approach~\cite{freysoldt2014first,van2004first}. Six $4d$/$5d$ transition metal defects are considered: Mo (4$d^5$5$s^1$), W (5$d^4$6$s^2$), Nb (\(4d^4 5s^1\)), Ta (\(5d^3 6s^2\)), Zr (\(4d^2 5s^2\)), and Hf (\(5d^2 6s^2\)).  A 2~$\times$~2~$\times~$1 hexagonal \CHR\ supercell with 120 atoms [see Fig.\ref{fig:structure}(b)] was built to host a single defect, and a $\Gamma$ point was used for Brillouin zone integration. 
The magnetic ground state of bulk \CHR\ is G-type antiferromagnetic. 
With the inclusion of the transition metal defects, the spin ordering of the bulk is typically preserved. However, multiple spin states, such as high-spin and low-spin configurations, can emerge at the defect site, depending on the competition between the crystal field splitting and the on-site exchange. To identify the most favorable spin state, we initialized the defect site with various local moments. The initial atomic positions of the defective supercell were randomly shifted and then relaxation was carried out until the Hellmann--Feynman forces on all atoms were reduced below 0.01 eV/\AA. Spin-orbit coupling is not included in our calculations. 

\subsection{Exchange energy}\label{method_sfe}  

To evaluate the influence of impurities on the N\'eel temperature of \CHR, we compute the exchange energy (hereafter denoted as $E_\text{ex}$), which represents the energy cost to invert the spin on a magnetic atom.~\cite{mu2013effect} This energy is directly associated with the strength of the exchange interaction between the central spin and its neighboring spins. For supercell calculations with impurity, we calculate the exchange energies on the cation impurity and nearby Cr atoms. Four neighboring Cr atoms are considered for the exchange energy calculation: one nearest Cr neighbor along the (0001) direction, and three next-to-nearest Cr neighbors in the buckled plane. [see Fig.~\ref{fig:structure}(b)] From our HSE calculations, the exchange energy for bulk \CHR\ is 178 meV. If the average value increases after doping, we expect an increase of the exchange interaction and the N\'eel temperature.

In classical mean field theory, the \NEL\ temperature is proportional to the exchange energy as:
\begin{equation}
T_{\mathrm{N}} = \frac{E_{\text{ex}}}{6k_{\mathrm{B}}}
\end{equation}

where \NELT\ represents the \NEL\ temperature, \(E_{\text{ex}}\) denotes the exchange energy, and \(k_{\mathrm{B}}\) is the Boltzmann constant. Hence, we can estimate the change in the \NEL\ temperature by comparing the average $E_\text{ex}$ of the defective supercell to that of the bulk. 
The average $E_\text{ex}$ is obtained by averaging $E_\text{ex}$ over all magnetic ions in the defective supercell.

\subsection{Defect formation energy}\label{formation_energy}

The stability of defects is assessed by calculating their formation energies $(E^f)$. The formation energy of a single cation substitutional defect $X$ in charge state $q$ is given by~\cite{freysoldt2014first}:
\begin{equation}
\begin{split}
E^f\left[X^q\right]=E_{\mathrm{tot}}\left[X^q\right]-E_{\mathrm{tot}}[\text { bulk }]-\mu_X \\
+\mu_\text{Cr}+q E_{\mathrm{F}}+\Delta^q.
\end{split}
\end{equation}

Here, $E_{\mathrm{tot}}[\text { bulk }]$ and $E_{\mathrm{tot}}\left[X^q\right]$ represent the total energies of a bulk supercell and a supercell with the defect, respectively. 
Different defect charge states were modeled by adding or removing electrons from the supercell using the \texttt{NELECT} parameter in VASP, with a uniform compensating background charge added.
Chemical potential for element $i$ is denoted as $\mu_i$. 
The Fermi level, or the electron's chemical potential, is denoted by $E_{\mathrm{F}}$ and is referenced to the valence-band maximum (VBM). The term $\Delta^q$ is a finite-size correction for the electrostatic interaction between a charged defect and its periodic images~\cite{freysoldt2009fullycorrection,freysoldt2011electrostatic}. This correction is calculated using an HSE dielectric tensor $\left(\epsilon^{||}=13.0, \epsilon^{\perp}=11.8\right)$, which aligns closely with the experimental dielectric constant~\cite{fang1963dielectric}.

The chemical potential of an atomic species, denoted as $\mu_i$, signifies the energy cost of atom exchange with a reservoir. It is given by $\mu_i=\mu_{\mathrm{ref}, i}+\Delta \mu_i$, where $\mu_{\mathrm{ref}, i}$ is the reference chemical potential of species $i$, and $\Delta{\mu_i}$ represents the atomic species' environmental abundance ~\cite{freysoldt2014first}. For Cr and O, the references are the energy of bulk Cr metal and the $\mathrm{O_2}$ molecule, respectively. In thermodynamic equilibrium, $\Delta{\mu_i}$ satisfies:
\begin{equation}
2\times\Delta \mu_{\mathrm{Cr}}+3\times\Delta \mu_{\mathrm{O}}=\Delta H_f(\mathrm{Cr_2O_3})  
\end{equation}
Here, $\Delta H_f(\mathrm{Cr_2O_3})$ is the formation enthalpy of \CHR. Our calculated  $\Delta H_f(\mathrm{Cr_2O_3})$ is 10.82 $\mathrm{eV}$, aligning well with the experimental value of 11.69 $\mathrm{eV}$~\cite{holzheid1995cr}. 
To prevent the formation of elemental phases, $\Delta \mu_i \leq 0$ . Under the Cr-rich conditions, $\Delta \mu_{\mathrm{Cr}}=0$, yielding $\Delta \mu_{\mathrm{O}}=\frac{1}{3}\Delta H_f(\mathrm{Cr_2O_3})$. Under O-rich conditions, $\Delta \mu_{\mathrm{O}}=0$, giving $\Delta \mu_{\mathrm{Cr}}=\frac{1}{2}\Delta H_f(\mathrm{Cr_2O_3})$. Chemical potentials of impurities are determined by their solubility-limiting phases, which are \( \mathrm{MoO_3} \) for \( \mathrm{Mo} \), \( \mathrm{Cr_2WO_6} \) for \( \mathrm{W} \), \( \mathrm{TaCrO_4} \) and \( \mathrm{Ta_2CrO_6} \) for \( \mathrm{Ta} \), \( \mathrm{NbCrO_4} \) for \( \mathrm{Nb} \), \( \mathrm{ZrO_2} \) for \( \mathrm{Zr} \), and \( \mathrm{HfO_2} \) for \( \mathrm{Hf} \). Under O-rich conditions, the limiting phases for Mo, W, Nb, Ta, Zr, and Hf are \(\ce{Cr2Mo3O12}\), \(\ce{Cr2WO6}\), \(\ce{NbCrO4}\), \(\ce{TaCrO4}\), \(\ce{ZrO2}\), and \(\ce{HfO2}\) respectively. As for the Cr-rich conditions, the limiting phases are bulk metallic Mo, bulk metallic W, \(\ce{NbCrO4}\), \(\ce{TaCrO4}\), \(\ce{ZrO2}\), and \(\ce{HfO2}\), respectively. 

Defect charge-state transition levels $\varepsilon\left(q / q^{\prime}\right)$ are defined as the Fermi-level position where the most stable charge state changes from $q$ to $q^{\prime}$. They are calculated as
\begin{equation}
\varepsilon\left(q / q^{\prime}\right)=\frac{E^f\left[X^q ; E_{\mathrm{F}}=0\right]-E^f\left[X^{q^{\prime}} ; E_{\mathrm{F}}=0\right]}{q^{\prime}-q} 
\end{equation}
where $E^f\left[X^q ; E_{\mathrm{F}}=0\right]$ is the formation energy of defect $X$ with charge $q$ when the Fermi level is at the VBM.

The binding energy of defect complex is also calculated to explore the stability of the complex once it is formed. Take $(\mathrm{Cr}_{i} - V_{\mathrm{Cr}})^n$ in charge state $n$ for example, the binding energy is defined in terms of formation energies,
\begin{equation}\label{eq:binding}
\begin{split}
E_{\text{bind}}[\left(\mathrm{Cr}_{i} - V_{\mathrm{Cr}}\right)^n] = & \ E^f\left(\mathrm{Cr}_{i}^{m}\right) \\
& + E^f\left(V_{\mathrm{Cr}}^{n-m}\right) - E^f\left[\left(\mathrm{Cr}_{i} - V_{\mathrm{Cr}}\right)^n\right],
\end{split}
\end{equation}
where $n$, $m$, and $n-m$ denote charge states of the complex, Cr$_i$ and $V_\text{Cr}$. Different $m$ are tested and the one with the lowest binding energy is adopted.  Charge neutrality relation is kept to evaluate the binding energy. By definition, a positive binding energy indicates a stable, bound defect complex.   

\subsection{Nudged elastic band calculation}
Migration barriers of Cr vacancy are calculated using the climbing-image
nudged elastic band (cNEB) method~\cite{henkelman2000climbing}. 
The initial and final structures correspond to a vacancy forming on two Cr sublattices connected by inversion symmetry. They are energetically degenerate but the net magnetizations of them are opposite. We divide the NEB paths into two segments, separated by an intermediate state of a split vacancy (denoted as V$^{s}_\textnormal{Cr}$). The split vacancy position depicts one Cr occupying the inversion center while two normal vacancies form on its nearby Cr neighbors along the $c$ axis. Due to the dual choice of Cr spins near the inversion center, two intermediate states with either spin choice are selected such that each segment of the NEB path is spin conserved. The atomic positions of two intermediate states are close, and we use a configurational-coordinate diagram to connect the two intermediate states. 

\section{Results and Discussions} \label{mainresult}

\subsection{\CHR\ bulk properties}

As mentioned in Sec.~\ref{sec:calc_detail}, the mixing parameter of 0.2 and screening parameter of 0.3 are chosen throughout DFT calculations, yielding good agreement with the measured bandgap, local moment on Cr and lattice parameters in bulk \CHR. A comparison of the lattice parameters, band gaps, and atomic positions between experiment and theory is shown in Table~\ref{tab:bulk_properties}. 
More details about the mixing parameter dependent bulk properties can be found in Sec. S1 of SM~\cite{supplemental}. 

\begin{figure}
\includegraphics[width=0.47\textwidth]{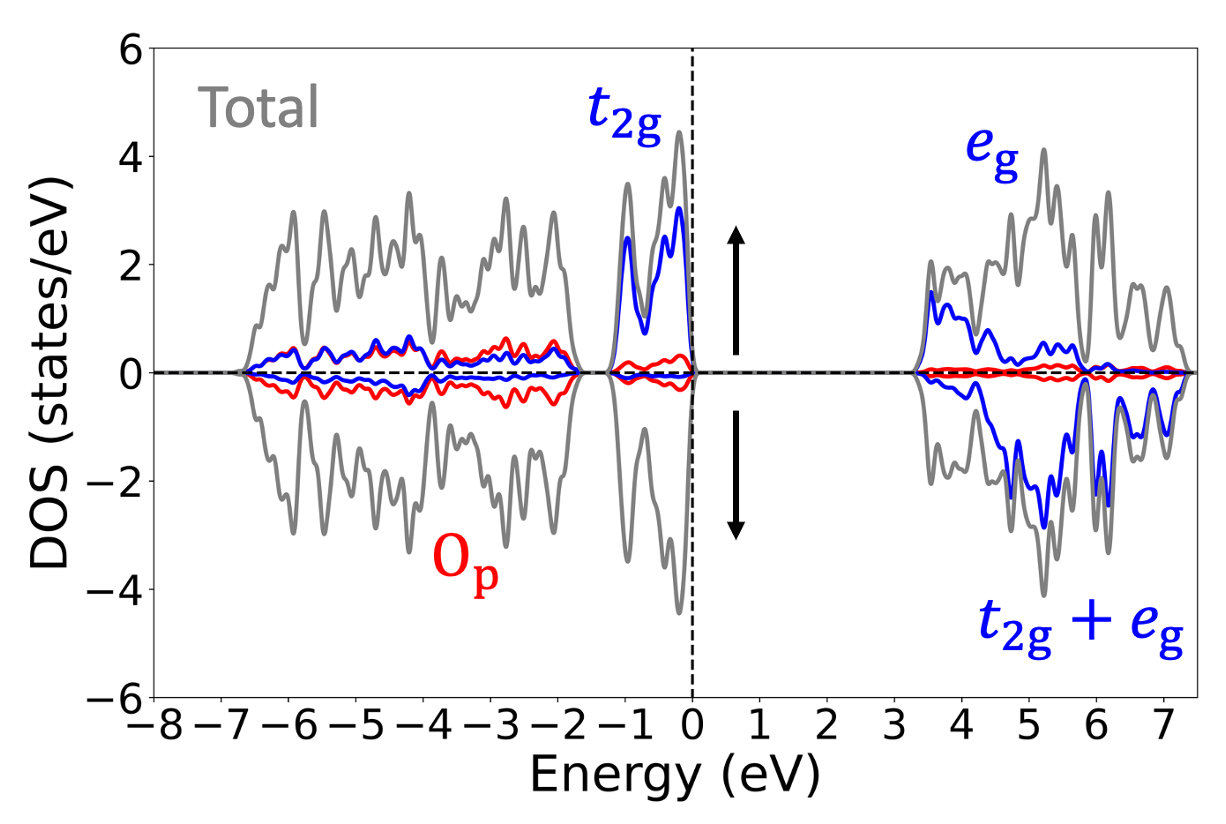}
\caption{\label{fig:formula_dos}Density of states per formula unit (f.u.) of \CHR\ bulk (grey lines) and partial density of states of $3d$ orbitals for one Cr with spin up local moment (blue lines) and O-$p$ orbitals (red lines). DOS in majority- and minority-spin channels are plotted as positive and negative, respectively. 
}
\end{figure}

Figure~\ref{fig:formula_dos} shows the calculated total and partial density of states (DOS) of \CHR\ using HSE, 
which agree well with previous works~\cite{shi2009magnetism}. 
Due to the quasi-octahedral crystal field, Cr 3$d$ orbitals are split into lower-lying $t_{2g}$ states and higher-lying $e_g$ states, which are labeled in Fig.~\ref{fig:formula_dos}. $t_{2g}$ orbitals are half-filled, which favors antiferromagnetic spin coupling for direct exchange interactions.~\cite{shi2009magnetism}   
Band structure of \CHR~is illustrated in the Fig. S2 of SM~\cite{supplemental}. An indirect band gap of 3.42 eV is found from calculation, which compares well with the measured indirect band gap of 3.40 eV~\cite{julkarnain2012optical,adler1968insulating}.

\begin{table}
\centering
\renewcommand{\arraystretch}{1.5} 
\caption{Lattice parameters (given in \AA), local moment ($\mu$, given in $\mu_B$) and bandgap of \CHR\ ($E_\text{g}$, given in eV).  Previous calculation and experiment results are compared.  Wyckoff positions for atoms in hexagonal \CHR\  are also given. }
\setlength{\tabcolsep}{10pt} 
\begin{tabular*}{\columnwidth}{@{\extracolsep{\fill}}lcccc} 
\hline
\multicolumn{3}{c}{} & This work & Expt \\
\hline
\multicolumn{3}{l}{$a\ (\mathrm{\AA})$} & 4.9668 & $4.9507^{\mathrm{a}}$\\
\multicolumn{3}{l}{$c\ (\mathrm{\AA})$} & 13.6472  & $13.5656^{\mathrm{a}}$ \\
\multicolumn{3}{l}{$\mu\ (\mu_B)$} & 2.86  & $2.48^{\mathrm{b}}$ \\
\multicolumn{3}{l}{$E_\text{g}$ (eV)} & 3.42  & $3.40^{\mathrm{c}}$ \\ 
\hline
\hline
Site & Position & $x$ & $y$ & $z$\\
\hline
Cr & 12$c$ & 0.0000 & 0.0000 & 0.6528 \\
O & 18$e$ & 0.3055 & 0.0000 & 0.2500 \\
\hline
\multicolumn{5}{l}{
$^{\mathrm{a}}$Reference\cite{finger1980crystalstructure};
$^{\mathrm{b}}$Reference\cite{brown2002determinationofmag};}\\
\multicolumn{5}{l}{$^{\mathrm{c}}$Reference\cite{crawford1964electricalCr2O3,adler1968insulating,zaanen1985bandgap}
}\\
\end{tabular*}
\label{tab:bulk_properties}
\end{table}

\subsection{Formation of intrinsic defects and complexes} \label{res:stability0}  
We first investigate intrinsic defects and their complexes in Cr$_2$O$_3$ for the following two purposes. 
First, it provides a complete picture of the native defect physics of Cr$_2$O$_3$, including the dominant defects under Cr-rich and O-rich conditions, their charge-state transition levels, the unique local structure configurations and the stability of defect complexes. 
Second, the study of intrinsic defects sheds light on possible Fermi-level pinning in \CHR, which is essential for determining the charge states of the transition-metal dopants discussed in the following section. 

The following intrinsic defects are considered: O vacancy ($\mathit{V}_{\mathrm{O}}$), O interstitial ($\mathrm{O}_{\mathit{i}}$), O substitution on Cr site ($\mathrm{O}_{\mathrm{Cr}}$), Cr vacancy ($\mathit{V}_{\mathrm{Cr}}$), Cr interstitial ($\mathrm{Cr}_{\mathit{i}}$), Cr substitution on O site (Cr$_\text{O}$), $\mathrm{Cr}_i\!-\!\mathit{V}_{\mathrm{O}}$ and $\mathrm{Cr}_i\!-\!\mathit{V}_{\mathrm{Cr}}$. 
Note that we were unable to converge the residual force for Cr$_\text{O}$, so it is not included in this study.  
Their formation energies under Cr-rich and O-rich conditions are shown in Fig.~\ref{fig:intrinsic_formation_energy}.

\subsubsection{Intrinsic defects}\label{sec:intrinsic}
\begin{figure}
\includegraphics[width=0.47\textwidth]{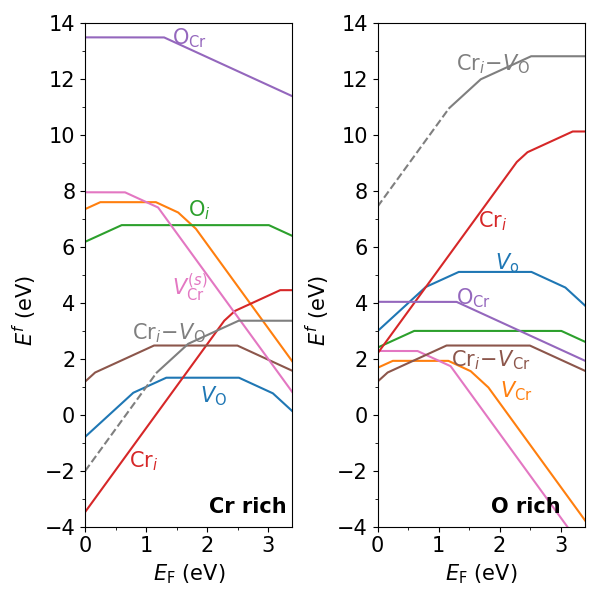}
\caption{\label{fig:intrinsic_formation_energy}Formation energy $E^f$ of intrinsic defects in \CHR\ as a function of Fermi level $E_{\mathrm{F}}$ under (a) Cr-rich and (b) O -rich conditions. }
\end{figure}

\paragraph{\textnormal{Cr} interstitial} 
There are two possible sites for cation interstitial: one octahedral site at the inversion center with six surrounding oxygen atoms (see Fig.~\ref{fig:structure}), and the other tetrahedral site with four oxygen vertices (see Fig.~S3 in the SM~\cite{supplemental}). 
Energetically, we found that Cr interstitial (Cr$_i$) strongly prefers the octahedral site at the inversion center and this is also true for all studied interstitial cation impurities. Therefore, we ignore the tetrahedral site when talking about the cation interstitial hereafter.   

The formation energy of interstitial chromium (Cr$_i$) under both O-rich and Cr-rich conditions is shown in Fig.~\ref{fig:intrinsic_formation_energy}. Once incorporated, Cr$_i$ in \CHR~acts as a dominant donor when the Fermi level lies deep within the band gap. The ($3+$/$2+$), ($2+$/$+$), and ($+$/$0$) charge-state transition levels are located 1.12, 0.94, and 0.20 eV below the CBM, respectively.
It is worth noting that while our calculated ($2+$/$+$) and ($+$/$0$) levels agree well with those reported in Ref.~\onlinecite{medasani2018first}, our ($3+$/$2+$) level is 0.64 eV higher. This discrepancy likely arises from differences in the computational methods: our calculations employ a hybrid functional, whereas Ref.~\onlinecite{medasani2018first} used GGA+$U$.
When the Fermi level is near midgap, Cr$_i$ favors the $3+$ charge state. Cr$_i$ is more readily incorporated under Cr-rich conditions due to its lower formation energy, which is 1.63 eV when the Fermi level is at midgap.
As a result, Cr$_i$ is expected to be the dominant defect in \CHR~under Cr-rich growth conditions, pushing the Fermi level toward the upper part of the band gap due to its donor character. 

It is worth noting that the Cr$^{3+}_i$ carries a local moment of 2.73 $\mu_B$.
The presence of a magnetic cation at the inversion center breaks the combined inversion--time-reversal symmetry (I$\otimes$T), where I and T denote inversion and time-reversal symmetry, respectively.   
The symmetry breaking induces a local structural relaxation involving Cr$_i$ and its six neighboring oxygen atoms. Cr$_i$ is slightly displaced from the inversion center, while the surrounding O atoms shift inward toward the interstitial site. As a result, the Cr$_i$-O bond lengths within the Cr$_i$O$_6$ octahedron are reduced to 1.99 \AA~ and 1.97 \AA, compared to the ideal bond length of 2.07 \AA~in the absence of symmetry breaking.   

\begin{figure}
\includegraphics[width=0.47\textwidth]{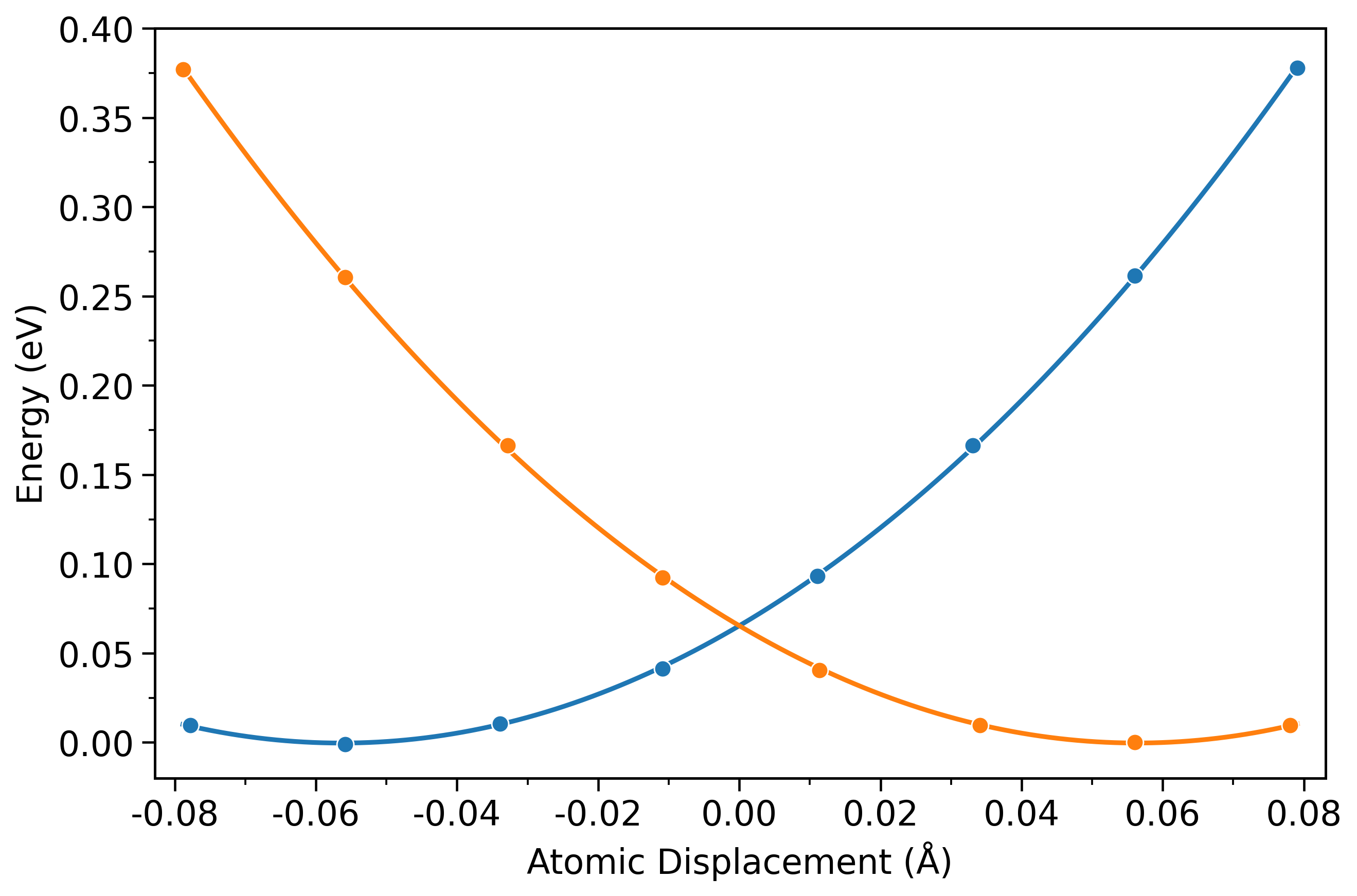}
\caption{\label{fig:Cr_i_configurational} Blue (orange) dots denote the calculated energy profile for an interstitial Cr atom displaced away from the inversion center (set as zero displacement) in the spin-down (spin-up) state. The blue (orange) lines are interpolated curves.}
\end{figure}

Despite the presence of a finite local moment on Cr$_i$, the exchange field at the inversion center is exactly zero by symmetry. This implies the existence of two degenerate Cr$_i$ configurations near the inversion center with opposite spin orientations. As expected, we identify both spin-reversed configurations, which are symmetry-equivalent. To evaluate the transition barrier between them, we computed the configurational coordinate diagram (see Fig.~\ref{fig:Cr_i_configurational}) and found a small migration barrier of 62 meV. 

\begin{figure}
\includegraphics[width=0.47\textwidth]{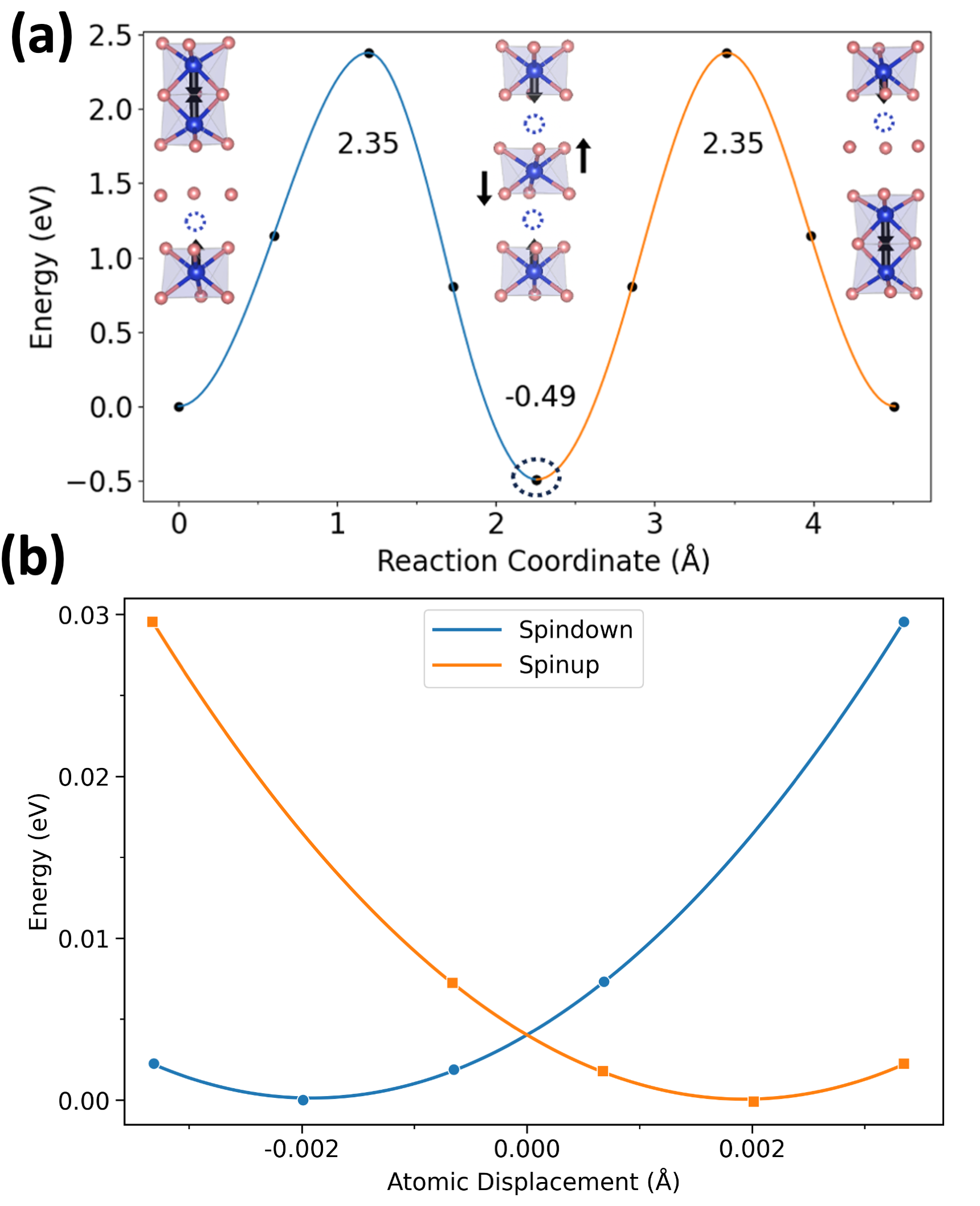} 
\caption{\label{fig:NEB} (a) Migration barrier of $V^{3-}_\text{Cr}$ from a normal Cr vacancy to a split vacancy and subsequently to another normal Cr vacancy with an opposite local moment. The spin-up and spin-down states are shown in orange and blue, respectively. The dashed circle indicates the transition between the spin-up and spin-down states near the inversion center. (b) Zoomed-in energy profile of $V^{(s)3-}_\text{Cr}$ as a function of the Cr displacement from the inversion center, which is set to zero displacement. }
\end{figure}

\paragraph{\textnormal{Cr} vacancy}
$V_\text{Cr}$ is a deep acceptor in \CHR~(see orange line in Fig.~\ref{fig:intrinsic_formation_energy}). The charge-state transition levels ($+$/0), (0/$-$),($-$/$2-$), ($2-$/$3-$) of $V_\text{Cr}$ are located at 0.25, 1.16, 1.52 and 1.82 eV above the VBM, respectively. When the Fermi level lies deep within the gap,  $V_\text{Cr}$ can occur in the $1-$, $2-$, and $3-$ charge states, acting as an acceptor. For all three negative charge states, the six O atoms around the vacancy relax outward by approximately 0.12 -- 0.14 \AA.  
Interestingly, we find that the conventional Cr vacancy ($V_\text{Cr}$) is only metastable. A more stable configuration emerges when a nearby Cr (fifth-nearest Cr neighbor to $V_\text{Cr}$) undergoes a significant relaxation, migrating along the hexagonal axis to an interstitial site (the inversion center, see Fig.~\ref{fig:structure}) in between, thereby forming a dumbbell structure along the hexagonal axis. 
Here we denote this Cr vacancy as split vacancy, i.e. $V^{(s)}_\text{Cr}$, which is equivalent to the formation of an interstitial Cr together with two Cr vacancies. 
The creation of the split vacancy is illustrated in Fig.~\ref{fig:NEB}. 
This energetically favorable Cr split-vacancy configuration has also been reported previously in Refs.~\onlinecite{lebreau2014structural,medasani2017vacancies}. Similar cation split-vacancy configuration has been identified in sapphire~\cite{hine2009supercell}, which has a crystal structure similar to that of \CHR.

As shown in Fig.\ref{fig:intrinsic_formation_energy}, only the (0/$-$) and ($-$/$3-$) charge-state transition levels of $V^{(s)}_\text{Cr}$ appear, located at 0.66 eV and 1.22 eV above the VBM, respectively. When the Fermi level lies deep within the band gap, $V^{(s)}_\text{Cr}$ adopts the $3-$ charge state. Formation of the $V^{(s)3-}_\text{Cr}$ configuration lowers the formation energy by 1.10 eV compared to the ordinary $V^{3-}_\text{Cr}$. However, realizing $V^{(s)3-}_\text{Cr}$ requires the migration of a nearby Cr atom to an interstitial site—an event that does not occur spontaneously due to a migration barrier. We find this barrier to be approximately 2.35 eV [see Fig.\ref{fig:NEB}(a)], indicating that the transition can be activated at high temperatures, allowing both vacancy configurations to form.
The NEB total-energy calculations indicate that $V^{(s)3-}_\text{Cr}$ is more stable than $V^{3-}_\text{Cr}$ by 0.49 eV. We note that this is different from the 1.10 eV difference in their formation energies (see Fig.~\ref{fig:intrinsic_formation_energy}). We find that the discrepancy arises from the charge correction which is included in formation energy calculations but ignored in NEB. Interestingly, the local charge distribution in $V^{3-}_\text{Cr}$ is around this normal vacancy, while for $V^{(s)3-}_\text{Cr}$, the charges tend to distribute closer to the two vacancy sites within this split vacancy.

We also find that the localized charges associated with the triply charged Cr vacancies ($V^{3-}_\text{Cr}$ and $V^{(s)3-}_\text{Cr}$) are spin-polarized, with a net magnetization of 3 $\mu_\text{B}$ for the whole supercell. 
Similar to Cr$^{3+}_i$, the interstitial Cr site within $V^{(s)3-}_\text{Cr}$ also carries a local moment of 2.7 $\mu_\text{B}$. Given that it occupies the inversion center with a minimal exchange field, its local moment could be easily flipped, together with a flipped total magnetization and a tiny local relaxation. Figure~\ref{fig:NEB} (b) shows the calculated energy profiles of both spin-up and spin-down states, as a function of reaction coordinate. The energy barrier to flip the local moment on the interstitial site of $V^{(s)3-}_\text{Cr}$ is 72 meV, comparable to the energy barrier (62 meV) for a spin flip in Cr$^{3+}_i$. The spin-flipped $V^{(s)3-}_\text{Cr}$ can further migrate along $c$ axis, converting to a normal Cr vacancy with an opposite spin. [see the other segment and structural illustration in Fig.~\ref{fig:NEB}(a)] 

Additionally, an in-plane split-vacancy configuration was also obtained, in which a vacancy forms at a Cr site located in the same plane as the inversion center, while the Cr atom related by inversion symmetry migrates to the center to form a dumbbell. The structure of this configuration is shown in Fig. S5 of the SM~\cite{supplemental}. However, its energy is at least 3.4 eV higher than that of the most stable split-vacancy configuration discussed earlier, and thus it can be safely excluded from further consideration.

\begin{figure}
\includegraphics[width=0.5\textwidth]{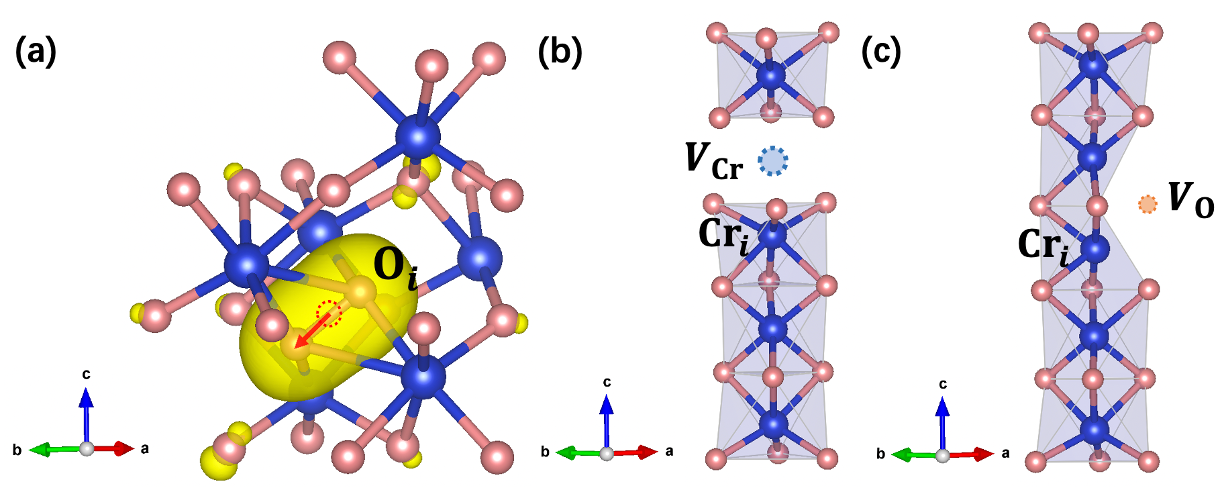}
\caption{(a) Local atomic environment around the \textit{neutral} O$^0_i$, featured by an interstitial O atom pushing a nearby O away from its ideal lattice site (dashed circle), forming a O--O dumbbell. 
The isosurface of charge density of O$^0_i$ is shown in yellow, and the isosurface level is 0.1 $e/{\text{\AA}^3}$. (b) $\mathrm{Cr}_i\!-\!\mathit{V}_{\mathrm{Cr}}$ complex with the Cr vacancy shown in black circle. (c) $\mathrm{Cr}_i\!-\!\mathit{V}_{\mathrm{O}}$ complex with the O vacancy shown in red circle.
}
\label{fig:defect_complex}
\end{figure}

In contrast to Cr interstitial, the formation energies of both split Cr vacancy ($V^{(s)}_\text{Cr}$) and normal chromium vacancy ($V_\text{Cr}$) are low under extremely O-rich conditions but increases significantly under Cr-rich conditions. Under practical experimental synthesis conditions which are typically intermediate between the Cr-rich and O-rich limits, Cr vacancies can appear in negatively charged states ($3-$, $2-$ for $V_\text{Cr}$ and $3-$ for $V^{(s)}_\text{Cr}$) in insulating samples and act as dominant compensators. 
Given their moderate formation energies, both $V_\text{Cr}$ and $V^{(s)}_\text{Cr}$ are thermodynamically accessible. 

Due to their acceptor nature and low formation energy under O-rich conditions, Cr vacancies are expected to dominate under synthesis environments involving O$_2$ gas flow. This can effectively shift the Fermi level toward the lower part of the band gap. However, the (0/$-$) transition level of the Cr vacancy is extremely deep, suggesting that Cr vacancies are unlikely to be the source of $p$-type conductivity observed in some \CHR~samples.~\cite{cao2006sol,cheng1996electrical}  

Experiments have observed a parasitic magnetization in Al(1.2\%)-doped \CHR~thin films and the parasitic magnetization can be controlled by oxygen flow condition during deposition, thereby by changing the oxygen content in the samples~\cite{nozaki2018manipulation}. It was demonstrated that more oxygen leads to a larger parasitic magnetization. The spin polarized Cr vacancies --  which dominates under O-rich conditions and could be easily aligned under magnetic field -- could be a possible reason for this parasitic magnetization. 

\paragraph{\textnormal{O} vacancy}
The oxygen vacancy $V_\mathrm{O}$  in \CHR~exhibits both donor and  acceptor levels within the gap. Four charge-state transition levels are identified: ($2+$/$+$), ($+$/$0$), ($0$/$-$), and ($-$/$2-$), locating at 0.79, 1.33, 2.50, and 3.10 eV above the VBM respectively. Each charge state is associated with different local lattice relaxations.
For $V^{2+}_\mathrm{O}$, the four nearby Cr atoms relax outward from the vacancy by 0.15$-$0.21 \AA. These  relaxations are reduced to 0.03$-$0.12 \AA~for $V^{+}_\mathrm{O}$ and to 0.01$-$0.02 \AA~for $V^{0}_\mathrm{O}$.  
In contrast, for the negative charge states of $V_\mathrm{O}$, the nearby Cr atoms move towards the vacancy by 0.02$-$0.09 \AA~for $V^{-}_\mathrm{O}$ and 0.05$-$0.12 \AA~for $V^{2-}_\mathrm{O}$. 
When Fermi level lies near mid-gap, especially between 1.33 and 2.50 eV above the VBM, $V_\mathrm{O}$ is most likely to form in the neutral charge state. Under Cr-rich conditions, the formation energy of $V^0_\mathrm{O}$ is relatively low, at 1.33 eV.  We further verified that $V^0_\mathrm{O}$ carries no spin moment and does not contribute to the global magnetization.

For a high-lying Fermi level, $V_\mathrm{O}$ will be in $1-$ or $2-$ charge state, depending on the position of the Fermi level. We note that those negatively charged states of $V_\mathrm{O}$ were missed by Ref.~\onlinecite{medasani2017vacancies}. On the other hand, Choi \emph{et al.} reported a negative charge state of $V_\text{O}$ in sapphire (Al$_2$O$_3$)~\cite{choi2013native}, which is iso-structural  to \CHR. This is consistent with our observation. 

\paragraph{\textnormal{O} interstitial}
We investigated three possible interstitial sites for oxygen in \CHR: (i) a tetrahedral site with four neighboring \textnormal{Cr} atoms (denoted as Tetra$_{i,\text{4Cr}}$), (ii) an octahedral site surrounded by six neighboring \textnormal{O} atoms (denoted as  Oct$_{i,\text{6O}}$) and (iii) a split interstitial configuration. 
These initial structures are illustrated in Fig.~S4 in SM~\cite{supplemental}. 
Among these,  O$_i$ prefers the tetrahedral site with four \textnormal{Cr} neighbors, although in the $1+$ and neutral charge states it forms a split interstitial . 
For example, in the case of neutral O$^0_i$, the interstitial O displaces a nearby lattice O atom, forming an O--O dumbbell structure, as shown in Fig.~S3.  The  O--O bond length for O$^+_i$ and O$^0_i$ are 1.44 and 1.42 \AA, respectively.  
For the negatively charged state O$^-_i$, the local structure  is similar, but the  O--O distance bond length increases to 1.91 \AA. 
the charge-state transition levels are calculated to be 0.61 eV above the VBM for the ($+$/0) level, and 0.40 eV below the CBM for the (0/$-$) level. 
Compared to other intrinsic defects, the formation energy of O$_i$ is significantly higher under Cr-rich conditions, making its incorporation unlikely. Even under O-rich conditions, where its formation energy is reduced, it remains substantially higher than that of Cr vacancies. 

\subsubsection{Defect complexes}
The migration barriers of Cr$_i$, O$_i$, $V_\text{O}$, $V_\text{Cr}$ in \CHR~have been previously calculated, revealing anisotropic migration behavior depending on the charge state of each defect.\cite{medasani2017vacancies, medasani2018first}
The reported migration barriers are relatively modest: the lowest migration barrier for $V_\text{O}$ is 1.22 eV, in-plane migration; For Cr$_i$, the minimum migration barrier is 1.86 eV along [221] direction; $V_\text{Cr}$ is the least mobile, with a barrier of 2.01 eV. 
Defect complexes involving these intrinsic defects -- such as  $V_\text{Cr}$-Cr$_i$ and Cr$_i$-$V_\text{O}$ -- can also form. In this section, we further calculate their formation energies and binding energies to evaluate their thermodynamic stability.  

The position of the Fermi level in a semiconductor with defects is governed by the condition of charge neutrality. This requires the total charge associated with all charged defects in insulating \CHR~to sum to zero. Under thermodynamic equilibrium, defect concentrations are determined by their formation energies.
Specifically, under Cr-rich conditions, the Cr interstitial ($\mathrm{Cr}_i$) is the dominant defect, whereas under O-rich conditions, the Cr vacancy ($V_{\mathrm{Cr}}$) prevails. 
Assuming that the Fermi level is pinned by $V_{\mathrm{Cr}}^{3-}$ and $\mathrm{Cr}_i^{3+}$, the formation energies of these key native defects suggest that the Fermi level lies near midgap, around 1.8 eV. 
The impact of impurities on this behavior will be discussed further in Sec.~\ref{result-form-sub}. It is important to note that electrically active impurities can shift the Fermi level position. 

\begin{figure}
\includegraphics[width=0.5\textwidth]{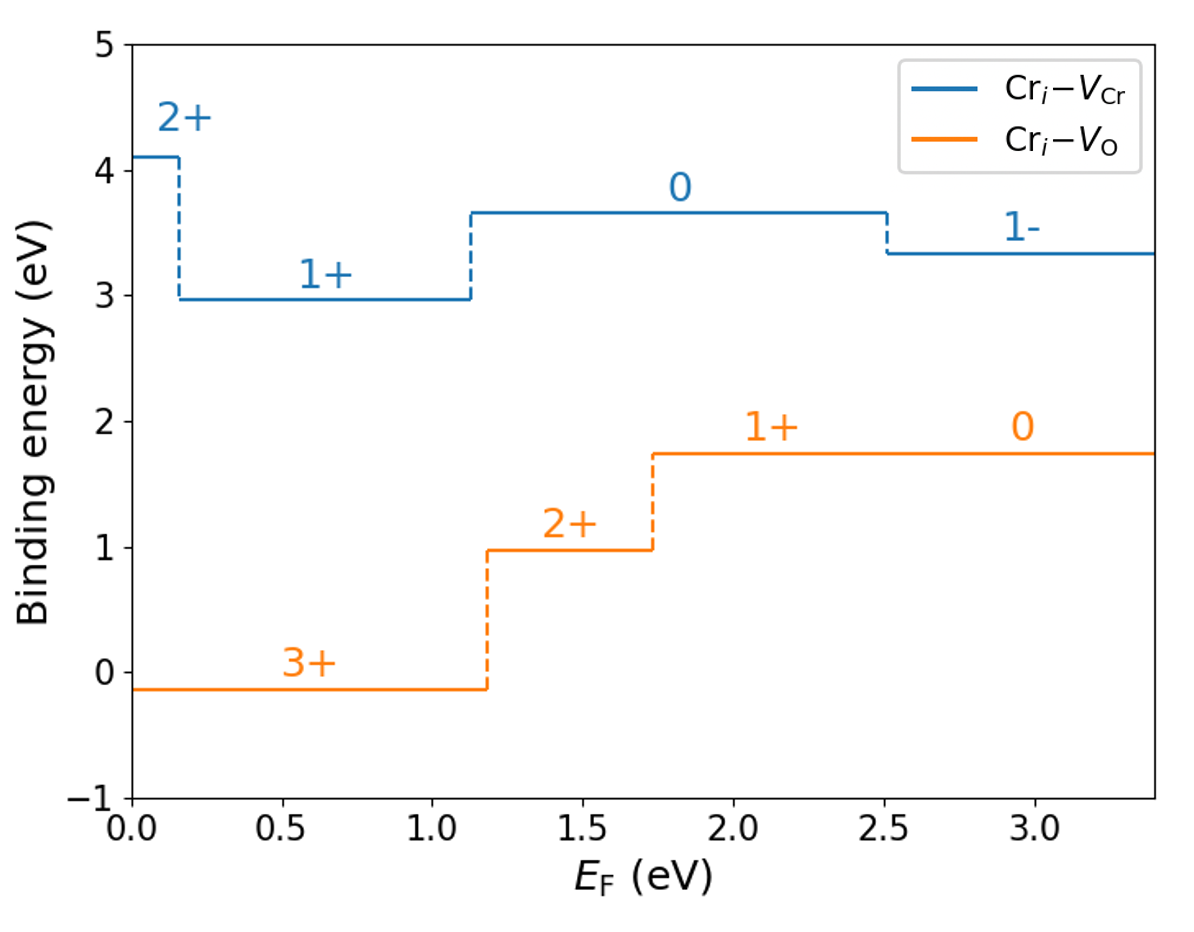}
\caption{Binding energies (eV) of $\mathrm{Cr}_i - V_\mathrm{Cr}$ and $\mathrm{Cr}_i - V_\mathrm{O}$ defect complex in different charge states.}
\label{fig:binding_energy}
\end{figure}

\paragraph{$\mathrm{Cr}_i - V_\mathrm{Cr}$}
A Frenkel defect forms when an atom or ion is displaced from its lattice site and relocates to a nearby interstitial position. We investigated two configurations of Cr-associated Frenkel defect, $\mathrm{Cr}_i-V_\mathrm{Cr}$: one with the vacancy and interstitial aligned along the hexagonal axis  (see Fig.~\ref{fig:defect_complex}b), and another with both located within the hexagonal plane. Our findings indicate that the former axial configuration is energetically favored in all charge states, by at least 2.5 eV. This defect complex exhibits multiple charge-state transition levels: ($+2$/$+$), ($+$/0), and (0/$-$), locating at \(0.16\), \(1.15\), and \(2.50\) eV above the VBM, respectively. 
When the Fermi level lies deep within the band gap, the $\mathrm{Cr}_i-V_\mathrm{Cr}$ complex is charge neutral, with a formation energy of 2.39 eV under Cr-rich conditions. Notably, the formation energy of $\mathrm{Cr}_i-V_\mathrm{Cr}$ under O-rich conditions is only slightly lower. 
To investigate the stability of the complex, we calculate its binding energy as a function of Fermi level using Eq.~\ref{eq:binding},  as shown in Fig.~\ref{fig:binding_energy}. This Frenkel defect is strongly bound, a binding energy greater than 3 eV across different charge states.  

\paragraph{$\mathrm{Cr}_i - V_\mathrm{O}$}
This defect complex consists of a Cr interstitial and a neighboring oxygen vacancy, as illustrated in Fig.~\ref{fig:defect_complex}(c). 
Three charge-state transition levels are identified within the band gap(3+/2+), (2+/+), and (+/0), located at 1.17, 1.70, and 2.52 eV above the VBM, respectively.
For deep Fermi levels, the $\mathrm{Cr}_i - V_\mathrm{O}$ complex is found in the neutral charge state.   
Under Cr-rich conditions, the formation energy of the  $\mathrm{Cr}_i - V_\mathrm{O}$ complex is moderately low, at approximately 2.50 eV, but it increases significantly under O-rich conditions. 
This indicates that the $\mathrm{Cr}_i - V_\mathrm{O}$ is easier to form under Cr-rich conditions, most likely in the neutral charge state.  
Although the $\mathrm{Cr}_i - V_\mathrm{O}$ complex is unstable in the $3+$ charge state due to a negative binding energy, it becomes stable in the $2+$, $+$, and neutral states (see the binding energy in Fig.~\ref{fig:binding_energy}).  Notably, in the neutral state, the binding energy is as high as 1.74 eV.   

\subsection{4$d$ and 5$d$ transition metal defects}\label{result-form-sub}

The intrinsic-defect results above are directly relevant to the transition-metal dopants considered below. Because the defect charge state depends on the Fermi-level position, the native defects in Cr$_2$O$_3$ can strongly influence which charge states of transition-metal dopant through Fermi-energy pinning.

Using LDA+$U$, Ref.~\onlinecite{mu2019influence} found that isovalent Mo (Mo$^0_\text{Cr}$) is favorable to increase the \NEL\ temperature by 8~\% per 1~\% Mo doping. Here we employ hybrid functional to conduct a systematic evaluation of the 4$d$ and 5$d$ transition metal defects, not only to address their impact on the \NEL\ temperature, but also investigate their likelihood of incorporation and clarify their charge states. Six 4$d$ and 5$d$ transition metal elements are studied: $X$ = Mo, W, Nb, Ta, Zr and Hf, and both interstitial  ($X_i$) and substitutional configurations ($X_\text{Cr}$) are considered.  

In addition, many of these defects are known to enhance the mechanical, thermal properties and functional properties of Cr$_2$O$_3$. For instance, Mo doping improves the passivity of Cr$_2$O$_3$~\cite{huang2023atomistic}, enhancing its corrosion resistance, while Zr doping increases hardness and provides excellent high-temperature thermal stability~\cite{mohammadtaheri2019investigation}, making it attractive for protective coating applications. Understanding the formation of these dopants is therefore technologically important for optimizing such functionalities.

\subsubsection{Substitutional defect: $X_\text{Cr}$}

\paragraph{$\textnormal{Mo}_\textnormal{Cr}$ $\textnormal{and}$ $\textnormal{W}_\textnormal{Cr}$}

\begin{figure}
\includegraphics[width=0.48\textwidth]{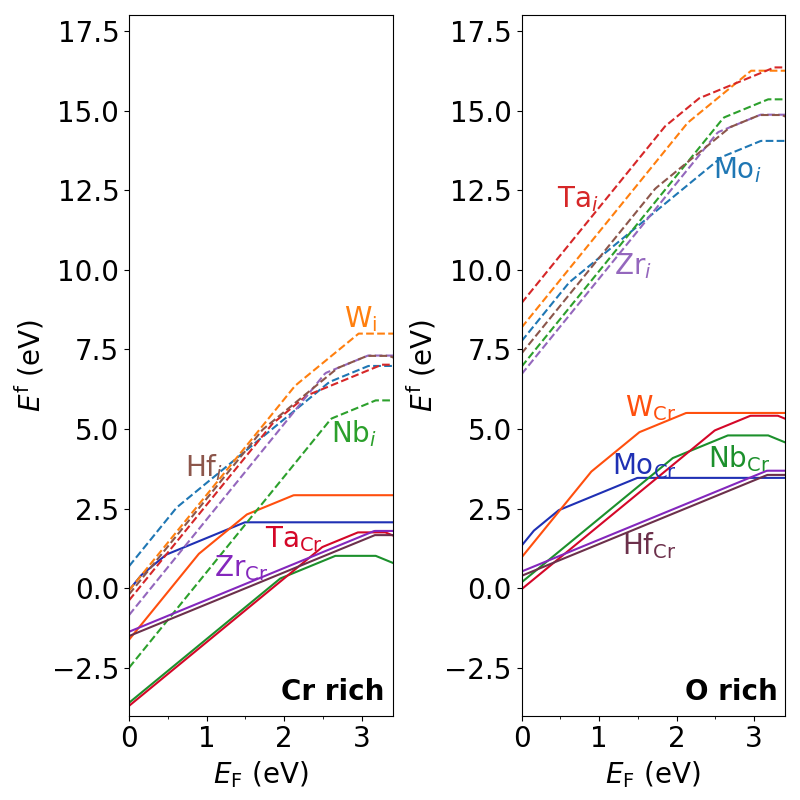}
\caption{\label{fig:TM_formation_energy}Formation energy \(E^f\) of 4\(d\) and 5\(d\) transition metal defects ($X_{i}$ and $X_\text{Cr}$, X = Mo, Nb, Zr, W, Ta, Hf) in \CHR\ as functions of the Fermi level \(E_{\mathrm{F}}\) under (a) O-rich and (b) Cr-rich conditions. $X_{i}$ represents interstitial defect, illustrated using dashed lines. $X_\text{Cr}$ denotes substitution defects on Cr site, illustrated using solid lines. 
}
\end{figure}

We first consider the substitutional defects $X_\text{Cr}$, and we start from Mo and W dopants. There is only one type of cation site in \CHR~(see Tab.~\ref{tab:bulk_properties} for its Wykoff position), giving one $X_\text{Cr}$. 
Isolated Mo (4$d^5$5$s^1$) and W (5$d^4$6$s^2$) atoms each have six valence electrons, similar to that of Cr (3$d^5$4$s^1$). One may expect that both Mo$_\text{Cr}$ and W$_\text{Cr}$ will naturally yield isovalent cations Mo$^{3+}$ and W$^{3+}$ as Cr$^{3+}$ in \CHR, i.e. $+3$ oxidation state. However, distinct from \CHR, Mo and W typically exhibit a $+6$ oxidation state, as in their stable oxide compounds MoO$_3$ and WO$_3$. This indicates that Mo$_\text{Cr}$ and W$_\text{Cr}$ in \CHR\ can act as donors. This agrees with what we found from the formation energy diagram for Mo$_\text{Cr}$ and W$_\text{Cr}$, as shown in Fig.~\ref{fig:TM_formation_energy}. 

For low Fermi levels in the gap, both Mo and W cations act as donors. $\textnormal{Mo}_\textnormal{Cr}$ and $\textnormal{W}_\textnormal{Cr}$ are in a 3+, 2+, or 1+ charge state depending on the position of the Fermi level. The (1+/0) charge state transition level occurs at 1.90 and 1.27 eV below the CBM for Mo$_\text{Cr}$ and W$_\text{Cr}$, respectively. Above the (1+/0) charge state transition level, both Mo$_\text{Cr}$ and W$_\text{Cr}$ will be in neutral charge state, i.e. isovalent to Cr$^{3+}$. For insulating \CHR\ thin films, with an $E_\text{F}$ pinned near the mid bandgap, Mo$_\text{Cr}$ will be in 0 or 1+ charge state and W$_\text{Cr}$ will be in 2+ or 1+ charge state. Only when the E$_\text{F}$ lies slightly higher than the midgap, W$_\text{Cr}$ could be in neutral charge state. We will evaluate the impact of these feasible charge states on the \NEL\ temperature of \CHR\ in Sec.~\ref{result:sfe}.  

The formation energy of $X_\text{Cr}$ under Cr-rich conditions is notably lower than that under O-rich conditions. While this may appear counterintuitive for a substitutional dopant on the Cr site, it arises from our choice to compute the formation energies for conditions corresponding to the solubility limit, which is set by the enthalpy of formation of the limiting phase and introduces a dependence of $\mu_\text{X}$ on $\mu_\text{O}$. The same trend has been reported for transition metal defects in $\beta$-Ga$_2$O$_3$~\cite{peelaers2016doping}. Even under O-rich conditions, when the Fermi level is located near the mid-gap, the Mo$^0_\text{Cr}$ is 3.46 eV, which is low enough to allow for the incorporation of Mo defects and the enhancement of corrosion resistance~\cite{huang2023atomistic}. On the other hand, the formation energy of W$_\text{Cr}$ is higher for high-lying Fermi level. The formation energy of W$^0_\text{Cr}$ is 0.86 eV higher than that of Mo$^0_\text{Cr}$ under Cr-rich conditions, indicating that the incorporation of W will be difficult. 

\paragraph{$\textnormal{Nb}_\textnormal{Cr}$ $\textnormal{and}$ $\textnormal{Ta}_\textnormal{Cr}$} 
Multiple charge states of Nb$_\text{Cr}$ and Ta$_\text{Cr}$($2+$, $1+$, $0$, $1-$) show up in their formation energy diagrams, as seen in the Fig.~\ref{fig:TM_formation_energy}. 
The ($+$/0) charge-state transition levels of Nb$_\text{Cr}$ and Ta$_\text{Cr}$ occur at 0.74 eV and 0.45 eV below the CBM, respectively. Below the ($+$/0) level, Nb and Ta act as donors. When $E_\text{F}$ lies in the mid-bandgap for insulating \CHR, both Nb$_\text{Cr}$ and Ta$_\text{Cr}$ will be a double donor, in $+2$ charge state. For high-lying $E_\text{F}$ that is close to the CBM, both Nb$_\text{Cr}$ and Ta$_\text{Cr}$ will act as acceptors, compensating the carriers. The  (0/$-$) charge-state transition levels occur at 0.22 eV and 0.09 eV below the CBM for Nb and Ta, respectively.    

Under O-rich conditions, the formation energies of Nb$^{2+}_\text{Cr}$ and Ta$^{2+}_\text{Cr}$ are 1.93 eV and 1.67 eV, respectively, when the Fermi level is pinned at the mid-gap. This is comparable to the formation energy of Mo$^0_\text{Cr}$. Similarly to Mo and W, the formation energies of Nb$_\text{Cr}$ and Ta$_\text{Cr}$ become much lower under Cr-rich conditions, due to the choice of solubility limit (see Sec.\ref{formation_energy} for details). Their negative formation energies under extreme Cr-rich conditions indicate that Nb and Ta are detrimental to the formation of bulk \CHR.  Successful incorporation of these defects should be achieved under conditions closer to O-rich, rather than Cr-rich, growth environments. 

\paragraph{$\textnormal{Hf}_\textnormal{Cr}$ $\textnormal{and}$ $\textnormal{Zr}_\textnormal{Cr}$} 
For a variety of $E_\text{F}$ positions in the bandgap, Hf$_\text{Cr}$ and Zr$_\text{Cr}$ are single donors in 1+ charge state. (see Fig.\ref{fig:TM_formation_energy}) 
This behavior contrasts with other cases where higher charge states are observed and this can be readily explained by the +4 natural oxidation state of these elements, as seen in stable oxide compounds such as HfO$_2$ and ZrO$_2$. 
Both of Hf$_\text{Cr}$ and Zr$_\text{Cr}$ are relatively shallow donors since their (+/0) charge-state transition levels occur 0.23 eV below the CBM. 

The formation energies of Hf$_\text{Cr}$ and Zr$_\text{Cr}$ are very similar for both O-rich and Cr-rich conditions.
Under O-rich conditions, Hf and Zr are the easiest to incorporate due to their lowest formation energies across all the investigated transition metal defects. Assuming a Fermi energy pinning at the midgap, the formation energies of Hf$^+_\text{Cr}$ and Zr$^+_\text{Cr}$ are 2.09 eV and 2.22 eV, respectively.  Notably, Zr incorporation is expected to enhance the mechanical properties of \CHR.~\cite{mohammadtaheri2019investigation}

\subsubsection{Exchange energy}\label{result:sfe} 

\begin{table*}[ht]
    \centering
    \renewcommand{\arraystretch}{1.5} 
    \setlength{\tabcolsep}{4pt} 
        \caption{Exchange energy on the defect site ($E_0$, meV) and on the first ($E_1$, meV) and second ($E_2$, meV) nearest Cr neighbors for different defect cases $X^q_\text{Cr}$ in charge state $q$. The enhancement in N\'eel temperature for 1~\% doping is denoted as $\eta$ ($\eta=\Delta T_{\mathrm{N}}/ T_\mathrm{N} $, \%). The reference exchange energy in bulk $\mathrm{Cr_2O_3}$ is 178 meV.}
    \begin{tabular}{l|c|c|c|c|c|c|c|c|c|c|c|c|c|c}
    \hline
      & $\mathrm{Mo_{Cr}^0}$ & $\mathrm{Mo_{Cr}^{1+}}$ & $\mathrm{W_{Cr}^0}$ & $\mathrm{W_{Cr}^{1+}}$ & $\mathrm{W_{Cr}^{1+}}$ & $\mathrm{Nb_{Cr}^0}$ & $\mathrm{Nb_{Cr}^{1+}}$ & $\mathrm{Nb_{Cr}^{2+}}$ & $\mathrm{Ta_{Cr}^0}$ & $\mathrm{Ta_{Cr}^{1+}}$ & $\mathrm{Ta_{Cr}^{2+}}$ & $\mathrm{Zr_{Cr}^{1+}}$ & $\mathrm{Hf_{Cr}^{1+}}$ & $\mathrm{Fe_{Cr}^0}$ \\
    \hline
    $E_0$  & 755 & 183 & 970 & 320 & 108 & 537 & 238 & -- & -- & 371 & -- & -- & -- & 102 \\
    \hline
    $E_1$  & 333 & 109 & 472 & 101 & 112 & 136 & 119 & 122 & 108 & 107 & 120 & 129 & 128 & 164 \\
    \hline
    $E_2$  & 279 & 183 & 346 & 168 & 221 & 349 & 305 & 141 & 307 & 162 & 144 & 144 & 145 & 166 \\
    \hline
    $\eta$  & 5.8 & -0.3 & 8.9 & 0.2 & -0.4 & 4.7 & 2.1 & 0.8 & 0.4 & 0.4 & -1.9 & -1.8 & -1.8 & -0.7\\
    \hline
\end{tabular}

    \label{tab:exchange_energies}
\end{table*}

To investigate the impact of transition metal defects on the N\'eel temperature of \CHR, we use the exchange energies on the defect and on nearby Cr atoms (defined in Sec.~\ref{method_sfe}) as indicators for the N\'eel temperature enhancement. If the exchange energies are increased after defect incorporation, we expect an enhancement in the N\'eel temperature of \CHR. Since the interstitial $4d$ and $5d$ transition metals ($X_i$) are difficult to be incorporated due to their high formation energies (see Sec.~\ref{result_xi}), we restrict ourselves to the discussion of the impact from substitutional defects ($X_\text{Cr}$). Since the substitutional defects exhibit multiple charge states for different positions of Fermi level, we focus on feasible charge states when $E_\text{F}$ lies near the midgap as in insulating samples. 

\begin{figure}
\includegraphics[width=0.47\textwidth]{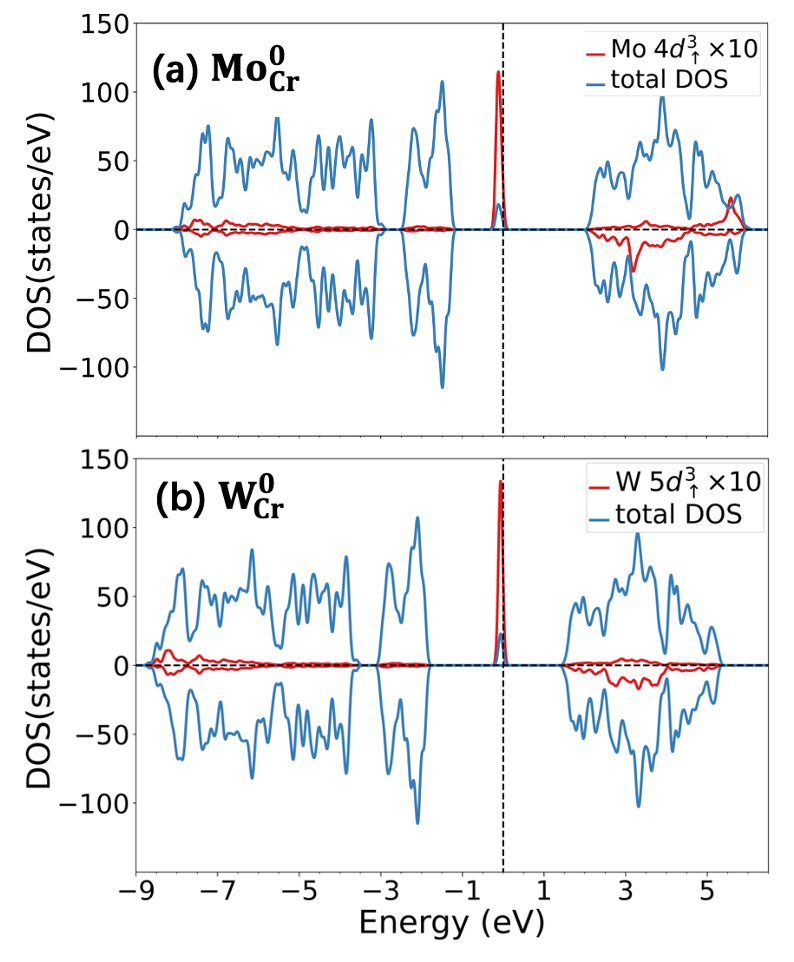}
\caption{\label{fig:Mo_Cr+W_Cr} (a) Total and Mo atom 4$d$ DOS for Mo$^0_\text{Cr}$ in \CHR.  (b) Total and W atom 5$d$ DOS for W$^0_\text{Cr}$ in \CHR.
}
\end{figure}

Table~\ref{tab:exchange_energies} summarizes the exchange energies on the defect site and also on the first and second nearest Cr neighbors for various defect cases. The exchange energy on Cr in bulk \CHR\ is 178 meV. To benchmark our hybrid functional calculations with previous DFT+$U$ studies~\cite{mu2013effect}, we calculated the exchange energy for a $3d$ iron impurity, and confirmed that it is detrimental to the $T_\text{N}$ enhancement (see Tab.~\ref{tab:exchange_energies}). This is consistent with Ref.~\onlinecite{mu2013effect}. 
Among the $4d$ and $5d$ impurity cases, the exchange energies on both \textit{neutral} Mo$^0_\text{Cr}$ and W$^0_\text{Cr}$ are drastically increased to 755 meV and 970 meV, respectively. 
Both Mo$^0_\text{Cr}$ and W$^0_\text{Cr}$ defects have the $d^3$ electron configuration, yielding a local moment of 3 $\mu_B$ (high spin). 
The $d^3$ electron configuration corresponds to half-filled t$_{2g}$ states, favoring \textit{antiferromagnetic} direct exchange~\cite{shi2009magnetism} and this is optimal for antiferromagnetic ordering. 
The remarkable energy enhancement for flipping the local moment of Mo$^0_\text{Cr}$ and W$^0_\text{Cr}$ can be understood from the electronic structure. As shown in Fig.~\ref{fig:Mo_Cr+W_Cr}, the impurity peaks for both Mo$^0_\text{Cr}$ and W$^0_\text{Cr}$ are located deep in the bandgap, mediating an enhanced direct exchange coupling between the spin on the defect and the spins on nearby Cr atoms. 
This reinforcement in the direct exchange coupling is facilitated by the reduced gap between filled (impurity) $d$ states and empty $d$ states for virtual electron hopping. 
The higher exchange energy of W$^0_\text{Cr}$ relative to Mo$^0_\text{Cr}$ is attributed to the fact that impurity peak of W$^0_\text{Cr}$ is closer to the mid gap and its excitation gap is smaller.
Additionally, the exchange energies on the Cr neighbors are increased for both Mo$^0_\text{Cr}$ and W$^0_\text{Cr}$. This enhancement arises from the in-gap impurity peaks associated with the $d^3$ electron configuration. Averaging the exchange energies, we can roughly estimate the enhancement in \NELT\ for both neutral defects: 1~\% W (Mo) substitution can increase $T_\text{N}$ by 8.9~\% (5.8~\%). 

If the Fermi level lies at midgap, charged defects W$^+_\text{Cr}$ and Mo$^+_\text{Cr}$ will emerge (see Fig.~\ref{fig:TM_formation_energy} and discussion of the charge-state transition levels in Sec.~\ref{result-form-sub}). 
We also investigate the exchange energies for W$^+_\text{Cr}$ and Mo$^+_\text{Cr}$, and there is no notable enhancement in the exchange energies (see Table.~\ref{tab:exchange_energies}). 
This can be explained by the $d^2$ configuration of both charged defects, which are not optimal for antiferromagnetic interactions. This finding also suggests us to facilitate the neutral charge state of Mo and W while the $1+$ charge state should be prevented. This can be achieved by adjusting the Fermi level just above the midgap to ensure their neutral charge state. From the study of intrinsic defect (Sec.~\ref{sec:intrinsic}), donor-like Cr$_i$ will become dominant under Cr-rich conditions. The synthesis under Cr-rich conditions would move the Fermi level higher in the band gap, but we should also keep a substantial excitation gap (between the filled impurity peak and the empty conduction band) to prevent degrading of the device.

\begin{figure}
\includegraphics[width=0.47\textwidth]{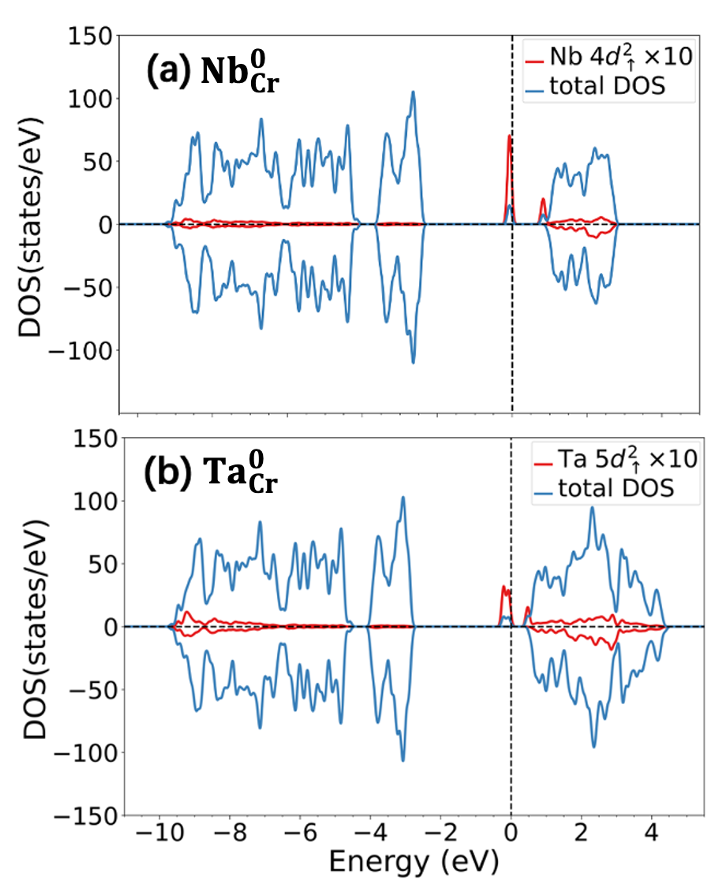}
\caption{\label{fig:Nb_Cr+Ta_Cr} (a) Total and Nb atom 4$d$ DOS for Nb$^0_\text{Cr}$ in \CHR. (b) Total and Ta atom 5$d$ DOS for Ta$^0_\text{Cr}$  in \CHR.
}
\end{figure}

Turning now to Nb$_\text{Cr}$ and Ta$_\text{Cr}$, Table~\ref{tab:exchange_energies} summarizes the exchange energies on the defect and nearby Cr atoms for $2+$, $1+$, and neutral charge states. The cases in which the exchange energies are notably enhanced are Nb$^0_\text{Cr}$ (by 4.7~\% per 1~\% doping) and Nb$^+_\text{Cr}$ (by 2.1~\% per 1~\% doping). As for Nb$^0_\text{Cr}$, the enhancement in the exchange energy is attributed to the enhanced exchange field on the defect site and also on the next-nearest Cr neighbors. The DOS of Nb$^0_\text{Cr}$ [see Fig.~\ref{fig:Nb_Cr+Ta_Cr} (a)] shows that the energy gap between the filled impurity levels and the unoccupied conduction bands is significantly reduced, which facilitates electron hopping. However, because the electronic occupancy deviates from the optimal half-filled $t_{2g}$ configuration, the antiferromagnetic coupling is weakened. As a result, the exchange energy is only moderately enhanced. 
On the other hand, while the DOS of Ta$^0_\text{Cr}$ is similar to that of Nb$^0_\text{Cr}$ [see Fig.~\ref{fig:Nb_Cr+Ta_Cr}(b)], Ta is found to be unfavorable for \NELT\ enhancement across all investigated charge states. 

\begin{figure}
\includegraphics[width=0.48\textwidth]{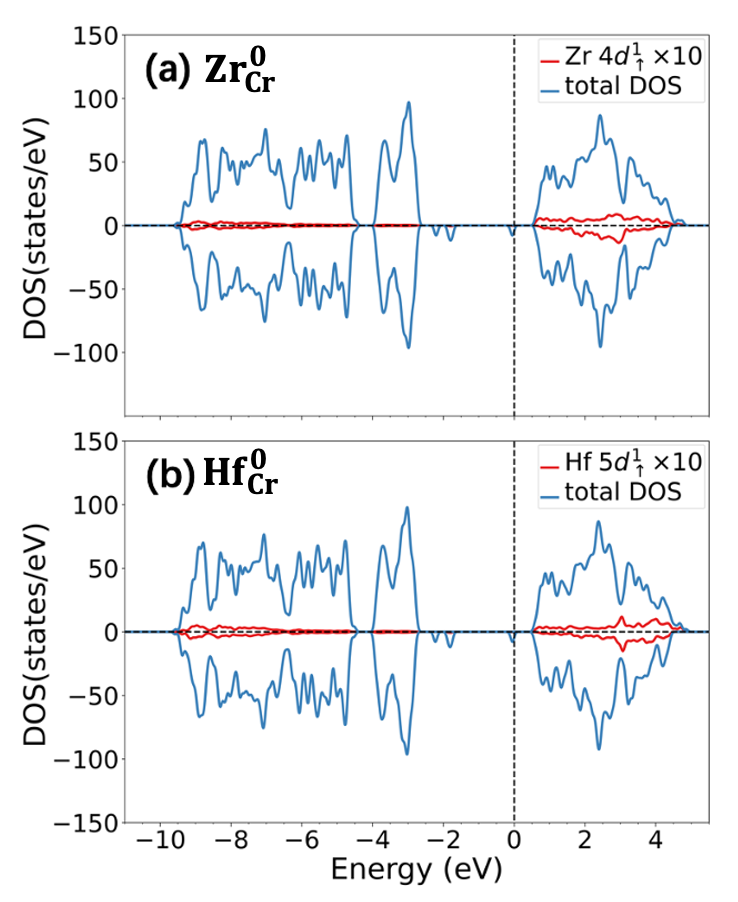}
\caption{\label{fig:Zr_Cr+Hf_Cr} (a) Total and Zr atom 4$d$ DOS for Zr$^0_\text{Cr}$ in \CHR. (b) Total and Hf atom 5$d$ DOS for Hf$^0_\text{Cr}$ in \CHR.
}
\end{figure}

The DOS of neutral Hf and Zr dopants are shown in Fig.~\ref{fig:Zr_Cr+Hf_Cr}. Although the ionic picture suggests one $d$ electron on the neutral cation, our calculations reveal nearly zero local magnetic moments due to low on-site exchange. For deep Fermi levels, Hf$^+_\text{Cr}$ and Zr$^+_\text{Cr}$ are the relevant charge states. However, since both Hf$^+_\text{Cr}$ and Zr$^+_\text{Cr}$ exhibit zero local moments, they do not contribute to any exchange energy directly.

Nevertheless, the dopant-induced local relaxations can modify the nearby Cr--Cr bond lengths, thereby altering the exchange interaction parameters between neighboring Cr atoms. To assess this effect, we calculated the exchange energies on nearby Cr atoms for Hf$^+_\text{Cr}$ and Zr$^+_\text{Cr}$. As shown in Table~\ref{tab:exchange_energies}, both dopants lead to slightly reduced exchange energies among neighboring Cr atoms, indicating that Hf and Zr are detrimental to the \NELT\ of Cr$_2$O$_3$.

\subsubsection{Interstitial defect: $X_\text{i}$}\label{result_xi} 
As mentioned in Sec.~\ref{sec:intrinsic}, the inversion center (previously denoted as $\mathrm{Octa}_{i, 6\mathrm{O}}$) is the most stable for Cr interstitials, and this is true for all 4$d$ and 5$d$ defects studied here. The interstitial defect $X_\text{i}$ are easier to form under Cr-rich conditions, however, their formation energies are still substantially higher than those of substitutional defect $X_\text{Cr}$ (see Fig.~\ref{fig:TM_formation_energy}).  Similar to Cr$_i$, multiple charge states of $X_\text{i}$ show up, and all $X_\text{i}$ are donors for deep Fermi level. All the charge state transition levels of $X_\text{i}$ are summarized in SM~\cite{supplemental}. 

Although these transition metal (TM) interstitials generally have higher formation energies compared to their substitutional counterparts, we still examine their exchange interactions. For deep Fermi levels, elements such as Ta, Zr, Hf, W, and Nb prefer the $3+$ charge state while Mo prefers $2+$ charge state. Interestingly, in Zr\(_i^{3+}\), Hf\(_i^{3+}\), and Nb\(_i^{3+}\) cases, we observe negative exchange energies, indicating that flipping the local magnetic moment on the first nearest neighbor of the defect atom leads to a lower total energy.    

Additionally, we find that certain charge states of interstitial defects can significantly increase $T_\text{N}$ of \CHR~compared to their substitutional counterparts. 
For instance, Hf\(_i^{2+}\) leads to a 27.2~\% increase of $T_\text{N}$ per 1~\% dopant, while Ta\(_i^{2+}\) doping yields a 14.1~\% increase per 1~\% dopant. 
The pronounced enhancement in the N\'eel temperature of Hf\(_i^{2+}\) is primarily attributed to a substantial increase in the exchange energy on second nearest neighboring (2NN) Cr atoms, from 178 and 979 meV. Moreover, an interstitial site at the inversion center is associated with six 2NNs, greatly surpassing three 2NNs as in the substitutional case, further amplifying the impact of 2NN on the N\'eel temperature.  Further investigating the DOS of Hf\(_i^{2+}\), we found that the substantial enhanced exchange energy on 2NN Cr atom comes from the electron redistribution of the in-gap states in two spin channels,  when the 2NN Cr spin is flipped.  Nevertheless, these interstitial TM defects are formidable to be incorporated due to their large formation energies, posting a great challenge to raise $T_\text{N}$ of \CHR. 

\section{Conclusions} 
\label{conc}
In summary, we conducted a systematic first-principles study of intrinsic and extrinsic (4$d$/5$d$ transition metal) defects in Cr$_2$O$_3$ to assess the likelihood of their incorporation and to identify strategies for enhancing its N\'eel temperature. Among the dopants examined, Mo and W substituting at the Cr site emerged as the most promising candidates. Both exhibit low formation energies under Cr-rich conditions and significantly enhance exchange interactions via impurity-mediated exchange interaction due to reduced hopping energy, particularly in their neutral charge states. Our estimates based on a mean-field model indicate that 1~\% Mo (W) doping can raise $T_\text{N}$ by approximately 5.8~\% (8.9~\%).  On the other hand, substitutional Nb, Ta, Zr, and Hf defects—despite their lower formation energies—are detrimental to $T_\text{N}$ enhancement. 
Additionally, interstitial transition metal defects are difficult to incorporate due to their high formation energies, although some of them at a certain charge state could potentially increase $T_\text{N}$.    

We also examined intrinsic defects, their complexes, and their electrical properties. Cr interstitials (Cr$_i$) and oxygen vacancies (V$_\mathrm{O}$) dominate under Cr-rich growth conditions and pin the Fermi level slightly above mid-gap, thereby stabilizing neutral Mo and W dopants, which are promising to enhance $T_\text{N}$. 
Conversely, Cr vacancies (V$_\mathrm{Cr}$) are dominant under O-rich conditions and tend to lower the Fermi level. A split Cr vacancy configuration is identified, exhibiting a slightly lower formation energy than the conventional vacancy. 
This split vacancy could be a possible source for the observed parasitic magnetization. A stable Frenkel pair, Cr$_i$--V$_\mathrm{Cr}$, with a large binding energy, is found to form readily under both Cr-rich and O-rich conditions and remains charge neutral when the Fermi level lies deep within the band gap. 
On the other hand, the Cr$_i$--V$_\mathrm{O}$ complex is stable only in the $2+$, $1+$, and neutral charge states, but its formation is likely only under Cr-rich conditions. 

The key ingredients for enhancing the N\'eel temperature of Cr$_2$O$_3$ via cation doping are summarized as follows:
\begin{enumerate}
    \item Cation defects should be easily incorporated, with low or moderate formation energies.
    \item The dopant should ideally exhibit half-filled electronic bands to maximize antiferromagnetic coupling.
    \item For semi-insulating films with a deep Fermi level, the desired charge state corresponding to half-filling should be accessible.
    \item The presence of in-gap impurity states is beneficial, as it can facilitate exchange interactions by lowering the electron hopping energy.
\end{enumerate}

Our study identifies Mo and W substitutional doping under Cr-rich conditions as a promising route to enhance the magnetic performance of Cr$_2$O$_3$ for spintronic applications. The detailed insights into defect energetics and magnetic interactions provide a solid theoretical foundation for future experimental optimization.

\begin{acknowledgments}
The authors acknowledge Tomohiro Nozaki for fruitful discussions and Zhi--Hao Wang for carefully reading the manuscript.
This work was funded in part by an ASPIRE grant from the VPR’s office at the University of South Carolina. S.M. acknowledges startup funding from the University of South Carolina. 
This work used the Expanse supercomputer at the San Diego Supercomputer Center through allocation PHY230073 and PHY230093 from the Advanced Cyberinfrastructure Coordination Ecosystem: Services \& Support (ACCESS) program, which is supported by the National Science Foundation (grant numbers \#2138259, \#2138286, \#2138307, \#2137603, and \#2138296).
The authors gratefully acknowledge the computational resources provided by the Hyperion high performance computing cluster at the University of South Carolina. We also acknowledge the technical assistance and resources provided by Research Computing at the University of South Carolina (RRID:SCR-027488).

\end{acknowledgments}

\end{document}